\documentclass[aps,prd,twocolumn,amsmath,amssymb,,groupedaddress,nofootinbib,numerical,superscriptaddress]{revtex4-1}

\usepackage{amsmath}
\usepackage{graphicx}
\usepackage{lipsum}
\usepackage{natbib}

\renewcommand{\a}{\hat{a}}
\renewcommand{\b}{\hat{b}}
\newcommand{\Dhr}{\hat{\Delta}^r} 
\newcommand{\Tmunu}{T_{\mu\nu}} 
\newcommand{\nmu}{n_{\mu}} 
\newcommand{\adot}{\dot{a}}
\newcommand{\addot}{\ddot{a}}

\begin{document}

\title{Fully relativistic nonlinear cosmological evolution in spherical symmetry using the BSSN formalism}

\author{J. Rekier${}^*$}
\affiliation{Namur Centre for Complex Systems (NaXys), University of Namur 
(UNamur), Belgium}
\email[]{jeremy.rekier@unamur.be}
\author{I. Cordero-Carri\'on}
\affiliation{Departamento de Matem\'atica Aplicada, Universidad de Valencia, E-46100 Burjassot, Spain}
\affiliation{Laboratoire Univers et Th\'eories (LUTh), Observatoire de 
Paris/CNRS/Universit\'e Paris Diderot, 5 place Jules Janssen, F-92190 Meudon, 
France}
\affiliation{Namur Centre for Complex Systems (NaXys), University of Namur 
(UNamur), Belgium}
\affiliation{IFPA, Dep.~AGO, Universite de Li\`ege, Bat B5, Sart Tilman B-4000 
Li\`ege 1, Belgium}
\author{A.F\"uzfa}
\affiliation{Namur Centre for Complex Systems (NaXys), University of Namur 
(UNamur), Belgium}

\date{\today}

\begin{abstract}
We present a fully relativistic numerical method for the study of cosmological problems using the Baumgarte-Shapiro-Shibata-Nakamura formalism on a dynamical Friedmann-Lema\^itre-Robertson-Walker background. This has many potential applications including the study of the growth of structures beyond the linear regime. We present one such application by reproducing the Lema\^itre-Tolman-Bondi solution for the collapse of pressureless matter with arbitrary lapse function. The regular and smooth numerical solution at the center of coordinates proceeds in a natural way by relying on the Partially Implicit Runge-Kutta algorithm described in Montero and Cordero-Carri\'on [arXiv:1211.5930]. We generalize the usual radiative outer boundary condition to the case of a dynamical background and show the stability and convergence properties of the method in the study of pure gauge dynamics on a de Sitter background. 
\end{abstract}


\maketitle

\section{Introduction}

Most cosmological models describe a spatially isotropic universe by using the Friedmann-Lema\^itre-Robertson-Walker (FLRW) metric~\cite{Lemaitre1933}. However, we do know that the Universe is neither perfectly homogeneous nor isotropic and cosmological inhomogeneities are unavoidable to account for numerous observations in cosmology, from the fluctuations in the CMB to the mechanism of structure formation. Many useful conclusions can already be drawn in spherical symmetry.

For what concerns structures formation, these studies are often performed either by assuming the Newtonian limit, in which the expansion of the Universe is considered as a small correction compared to the local gravitational fields and is adequately described by means of Newtonian dynamics, or the so called \emph{top-hat} approximation, in which the local inhomogeneity has a density profile in the shape of a step function. It is then assumed that spacetime inside the spherical object is described by a separated FLRW solution~\cite{Padmanabhan1993}. The first procedure lacks the merit of providing a fully general relativistic treatment of the problem. The second relies on debatable assumptions which would need sound reasons to be trusted throughout the whole spherical collapse process yet to be provided. 

Some simple cases escape these considerations by resorting to using the Lema\^itre-Tolman-Bondi (LTB) solution~\cite{Tolman1934,Bondi1947}. This metric, in its original form, only accounts for the collapse of pressureless matter (dust). It is worth noting that recent works extended this solution to the case of a general fluid~\cite{Lasky2006,Lasky2007}. In these, the Arnowitt-Deser-Misner (ADM)~\cite{ADM1962} formulation of general relativity is used to write equations for the LTB-like degrees of freedom in the form of an initial value problem using a single coordinate chart. As the authors acknowledge, the resulting equations are difficult to solve and need a numerical treatment. 

On the other hand, progress in the field of numerical relativity over the last two decades has allowed to solve many problems on asymptotically flat spacetimes with great accuracy. In these studies, assuming spherical symmetry reduces the number of spatial dimensions along with the cost of numerical computation. When dealing with spherical coordinates one has to take into account the $1/r^p$ terms close to $r=0$. The partially implicit Runge-Kutta (PIRK) methods presented in Refs.~\cite{Cordero-Carrion2012,Cordero2014} are an easy way to solve the problem of instabilities without the need to include any further regularization technique. These methods were proved helpful in the study of black hole evolution with the puncture gauge and spherical collapse of Tolman-Openheimer-Volkov solution solving Eintein's equations in the BSSN formalism~\cite{Nakamura1987,Shibata1995,Baumgarte1998}. Another way of regularizing the solution described for the ADM formalism in Ref.~\cite{Alcubierre2005} involves the inclusion of auxiliary variables and their corresponding evolution equations, which has been applied later successfully to the equations in BSSN formalism in Ref.~\cite{Alcubierre2011}. This strategy is more complex and demands more computer resources than the PIRK methods.


The BSSN formalism has already been applied on an expanding background by Shibata in Ref.~\cite{Shibata1999} for the study of primordial black holes (PBH) formation. This work involves the inclusion of a scalar field of matter and shows how the formation of a PBH depends on the initial energy profile. We wish to give further applications of this formalism by showing how it is also suitable to the study of the nonlinear growth of spherical structures in the Universe at later time. The present paper deals with dust matter and recovers the well-known LTB solution when the geodesic slicing is employed. The study of the nonlinear cosmological collapse is most interesting in presence of more general forms of matter. However, the case where matter is described by a quintessential scalar field is ongoing at the time of the writing of this manuscript and shall be published in an forthcoming paper. As a first step towards this, we present the case where the evolution of dust with a lapse function different than unity. In this case, the solution is nonanalytic and the evolution of matter is solved by employing the full hydrodynamical conservation equations. This allows us to test the method in a simple case and can be readily generalised to other kinds of matter. 

The formalism used is developed in some details in Sec.~\ref{section:formalism}. The specifications of the code are presented in Sec.~\ref{section:implementation}, including description of the Numerics, evolution scheme and boundary conditions. The numerical analysis of the code is presented in some details in Sect.~\ref{section:gaugedyn}, where it is applied to the study of pure gauge dynamics. In Sect.~5 we show an application to the collapse of pressure-less matter and we compare it with the LTB solution. We also give a generalisation of this solution to the case of non-unitary lapse. Geometrical units are used in which $G=c=M_{\odot}=1$. $\partial_i$ denotes partial derivative with respect to the corresponding variable.

\section{Formalism}
\label{section:formalism}

Under the assumption of spherical symmetry, the metric line element can be defined as 
\begin{equation}
ds^2=-(\alpha-\beta^{2})dt^2+2\beta drdt+\psi^4a^2(t)(\a~ dr^2+\b~ r^2d\Omega^2),
\label{eq:ansatz}
\end{equation}
where $\alpha$ is the lapse function, $\beta$ is the radial component of the shift vector, $\hat{a}$ and $\hat{b}$ are the nonzero components of the conformal 3-metric, the conformal factor is written as $\psi \sqrt{a}$, and all the variables are functions of $t$ and $r$. The choice of the lapse function and the shift vector defines the foliation of spacetime in spatial hypersurfaces. Following the strategy in Ref.~\cite{Shibata1999}, we factor out the cosmological scale factor $a(t)$, from the spatial 3-metric.

The Einstein equations describe the dynamics of spacetime. In all formulations, these equations are split into two groups: the constraint equations and the evolution equations. The BSSN scheme has proved to be very stable and is one of the most used formulations in numerical simulations. The dynamical variables in spherical coordinates have been listed in Ref.~\cite{Alcubierre2011}. We recall them here using slightly different notations. They consist in the lapse $\alpha$ and the shift $\beta$, the (redefined) conformal factor $\psi$ and the conformal metric functions $\a$ and $\b$. One has to add to this list the components of the extrinsic curvature, $K_{ij}:=-\frac{1}{2}\pounds_{\bf n}\gamma_{ij}$, which can be decomposed in trace $K$ and conformally scaled trace-free part $\hat{A}_{ij}$ as
\begin{equation}
K_{ij} = \frac{1}{3} \gamma_{ij} K + \psi^4 a^2 \hat{A}_{ij}.
\end{equation}
In spherical symmetry, $\hat{A}_{ij}$ has only two nonzero components, $A_a:=\hat{A}^r_r$ and $A_b:=\hat{A}^\theta_\theta$. Since $\hat{A}_{ij}$ must be traceless, one further has $A_a+2A_b=0$.

The great stability of the BSSN scheme is due to the addition of the auxiliary 3-vector $\Delta^i$. This vector has only one component in spherical symmetry (see e.g. Ref.~\cite{Alcubierre2011}):
\begin{equation}
\Dhr = \frac{1}{\a}\left[\frac{\partial_r\a}{2\a}-\frac{\partial_r\b}{\b}-\frac{2}{r}\left(1-\frac{\a}{\b}\right)\right].
\end{equation}
This is the last of our dynamical variables.

In what follows, we limit ourselves to the case with zero shift, $\beta = 0$. There is no formal difficulty in choosing a different gauge, but for the present purpose this choice allows more straightforward comparison with other cosmological evolutions.

The Einstein's equations are sourced by the energy content of the spacetime described by the energy-momentum tensor, $\Tmunu$. The energy source terms as seen by an Eulerian observer with 4-velocity $\nmu := (-\alpha,0)$ are:
\begin{align}
	E&=n_{\mu}n_{\nu}T^{\mu\nu}, \nonumber \\
	j_{i}&=-\gamma_{i\mu}n_{\nu}T^{\mu\nu}, \nonumber \\
	S_{ij}&=\gamma_{i\mu}\gamma_{j\nu}T^{\mu\nu},
\end{align}
where $E$, $j_i$ and $S_{ij}$ are the energy density, momentum density and stress energy tensor, respectively. Spherical symmetry reduces the number of such independent quantities to $E$, $j^r := \gamma^{ri}j_i$, $S_a := S^r_r$ and $S_b := S^\theta_\theta$.

The evolution equations for the dynamical variables are~\cite{Alcubierre2011}: 
\begin{align}
\partial_{t} \a &= -2\alpha \a A_{a},\nonumber\\
\partial_{t} \b &=-2\alpha \b A_{b}, \nonumber\\
\partial_{t} \psi &=-\frac{1}{6}\alpha \psi K - \frac{1}{2}\frac{\dot{a}}{a}\psi, \nonumber\\
\partial_{t} K  & = - \nabla^{2}\alpha +
\alpha(A_{a}^{2} + 2A_{b}^{2} + \frac{1}{3}K^{2}) \nonumber \\
    & + 4\pi\alpha(E+S_{a}+2S_{b}),\nonumber\\
\partial_{t} A_{a} & = - \left(\nabla^{r}\nabla_{r}\alpha - \frac{1}{3}\nabla^{2}\alpha\right)
+ \alpha\left(R^{r}_{r} - \frac{1}{3}R\right) \nonumber \\
   & + \alpha K A_{a} - \frac{16\pi}{3}\alpha(S_a - S_b)\nonumber\\
\partial_{t}\hat {\Delta}^{r} & = - \frac{2}{\a}(A_{a}\partial_{r}\alpha + \alpha\partial_{r}A_{a}) + 2\alpha\left(A_{a}\hat{\Delta}^{r} - \frac{2}{r\b}(A_{a}-A_{b})\right)\nonumber \\
    &+ \frac{\xi \alpha}{\a} \left[\partial_{r}A_{a} - \frac{2}{3}\partial_{r}K + 6A_{a}\frac{\partial_{r}\psi}{\psi} \right.\nonumber\\
    & \left.+(A_{a}-A_{b})\left(\frac{2}{r}+\frac{\partial_{r}\b}{\b}\right) - 8\pi j_{r} \right],
\end{align}
together with the evolution of the scale factor $a(t)$, and the asymptotic value of the lapse function $\alpha_{\tt bkg}$ (see next section for more details). Following Ref.~\cite{Montero2012}, we specify $\xi = 2$ to ensure strong hyperbolicity of the BSSN equations. The above feature the radial component of the Ricci tensor $R^r_r$ and its trace $R$, as well as the quantities $\nabla^{r}\nabla_{r}\alpha$ and $\nabla^{2}\alpha$. Their complete expressions in terms of the dynamical variables are detailed in Appendix~\ref{sect:algexpr}.

The Hamiltonian and momentum constraint equations have to be fulfilled on each spatial hypersurface and are given by
\begin{align}
  \mathcal{H} & \equiv R - (A^{2}_{a} + 2A^{2}_{b})
+ \frac{2}{3}K^{2}-16\pi E = 0, \\
  \mathcal{M}^r &\equiv \partial_{r}A_{a} 
- \frac{2}{3}\partial_{r}K + 6A_{a}\frac{\partial_{r}\psi}{\psi} \nonumber\\
  &+(A_a -A_b)\left(\frac{2}{r} + \frac{\partial_r \b}{\b}\right) - 8\pi j_r = 0.
\end{align}

\section{Implementation}
\label{section:implementation}

\subsection{Numerics}
The radial dimension is approximated by a uniformally discretised cell-centred grid, and radial derivatives are computed with a fourth-order finite difference scheme. We use fourth-order Kreiss-Oliger dissipation~\cite{Kreiss-Oliger1973}. The evolution equations are solved in time with the PIRK methods~\cite{Cordero-Carrion2012,Cordero2014}, and the applications to the evolution of BSSN variables has been described in~\cite{Montero2012}. We only present here a short summary. The method involves a splitting of the evolution equations for the dynamical variables as follows:
\begin{equation}
\left\{\begin{array}{rcl}
\partial_t u & = & \mathcal{L}_1(u,v),\\
\partial_t v & = & \mathcal{L}_2(u) + \mathcal{L}_3(u,v).
\end{array}
\right .
\end{equation}
In a first step of the evolution, $u$ is numerically evolved in an explicit way. The result is then used to evolve $v$ partially implicitly, making use of updated values of $u$ in the evaluation of the $\mathcal{L}_2$ operator.\footnote{The discrete evolution scheme used in the present paper is a second-order PIRK method, and involves a two-stage method described in detail in Ref.~\cite{Cordero-Carrion2012,Cordero2014,Montero2012}.}

The cosmological variables $a$ and the lapse of the background metric $\alpha_{\tt bkg}$ are first evolved explicitly along $\a$, $\b$, $\psi$ and $\alpha$. The updated values are then used to evolve $\dot{a}$, $K$ and $A_a$ partially implicitly. Finally, the update values are used to evolve $\hat{\Delta}^r$ partially implicitly.

\subsection{Cosmological evolution and boundary conditions}
Regularity of the dynamical variables close to the origin is enforced, in part, by specifying their parity across the origin. To achieve this in time, a few virtual points of negative radius are added to the numerical grid.

The considered spacetimes are not asymptotically flat. Instead, they tend to the cosmological FLRW solution. It is important to note that this work takes the cosmological solution as a homogeneous background on which the local inhomogeneous fields have no influence. 

The Friedmann and acceleration equations in the zero-shift gauge with arbitrary lapse are given by
\begin{align}
\frac{1}{\alpha_{\tt bkg}^2}\left(\frac{\adot}{a}\right)^2 &= \frac{8\pi}{3} \rho_{\tt bkg},
\label{eq:Friedmann}\\
\frac{1}{\alpha_{\tt bkg}^2}\frac{\addot}{a}-\frac{\dot{a}}{a}\frac{\dot{\alpha}_{\tt bkg}}{\alpha_{\tt bkg}}&= -\frac{8\pi}{6} (\rho_{\tt bkg}+3p_{\tt bkg}),
\label{eq:acc}
\end{align}
where $\rho_{\tt bkg}$ and $p_{\tt bkg}$ denote the homogeneous background energy density and pressure.

We impose radiative boundary conditions at the outer boundary
\begin{equation}
\partial_t f = \partial_t f_{\tt bkg} - v \, \partial_r f - \frac{v}{r} (f-f_{\tt bkg}),
\end{equation}
where $v$ is the speed of propagation of the variable $f$ on the grid. This is inferred by considering the characteristic structure of the variables of the evolution system of equations. In the above $f_{\tt bkg}=f_{\tt bkg}(t)$ denotes the spatially homogeneous asymptotic cosmological value of the variable $f$ and $\partial_t f_{\tt bkg}$ its first time derivative. These expressions can be read from their asymptotic values
\begin{align}
\a(t,r), \b(t,r), \psi(t,r) &\rightarrow 1,\nonumber\\
\alpha(t,r) &\rightarrow \alpha_{\tt bkg}(t).\nonumber\\
\end{align}
From the definition of the extrinsic curvature tensor, one has
\begin{align}
A_a(t,r), A_b(t,r)&\rightarrow 0,\nonumber\\
K(t,r) & \rightarrow -3 \frac{1}{\alpha_{\tt bkg}}\frac{\adot}{a},\nonumber\\
\end{align}
and from the definition of the $\Dhr$ variable one further has that
\begin{equation}
\Dhr \rightarrow 0.
\end{equation}
Other outer boundary conditions have been already presented in \cite{Zilhao2012} for a dynamical simulation of a spacetime in a cosmological background of FLRW.

\section{Code validation: Pure Gauge Dynamics}
\label{section:gaugedyn}
\subsection{Equations}

In order to validate our numerical code, we consider here pure gauge dynamics on a dynamical de Sitter background. This is the solution for a universe filled with a constant homogeneous vacuum energy density with equation of state $p_{\tt bkg} = - \rho_{\tt bkg}$. This is equivalent to adding a cosmological constant $\Lambda$ to the Einstein equations such that $\Lambda = 8\pi\rho_{\tt bkg}$.

We study the dynamical evolution of a Gaussian gauge pulse in the manner of Refs.~\cite{Alcubierre2011,Montero2012}. The only difference is that in our case the lapse function is ``perturbed'' around its cosmological (nonasympotically flat) value. 

Initially, we set
\begin{equation}
\alpha(t=0)  = \alpha^0_{\tt bkg}+\frac{\alpha_0r^2}{1+r^2}\left[e^{-(r-r_0)^2}+e^{-(r+r_0)^2}\right],
\end{equation}
where $\alpha^0_{\tt bkg}=\alpha_{\tt bkg}(t=0)$ and $\alpha_0$ is a constant which sets the amplitude of the Gaussian perturbation. Setting $\alpha^0_{\tt bkg}=1$ in~(\ref{eq:Friedmann}) allows us to define the initial Hubble factor $\displaystyle H_0 := \frac{\dot{a}(t=0)}{a(t=0)} = \frac{\dot{a}_0}{a_0}$. Note that, since the energy density remains constant, one has 
\begin{equation}
\frac{1}{\alpha_{\tt bkg}}\frac{\dot{a}}{a} = H_0, \forall t.
\label{eq:const_H0}
\end{equation}
The energy component is at all times equal to the constant homogeneous cosmological density:
\begin{equation}
E = \rho_{\tt bkg} = \frac{3}{8\pi}H_0^2.
\end{equation}

All the dynamical variables must fulfill both the Hamiltonian and the momentum constraints. By comparison with the homogeneous cosmological case, we set
\begin{equation}
\a(t=0) = \b(t=0) = \psi(t=0) = 1.
\end{equation}
These assumptions and the fact that $A_b=-\frac{1}{2}A_a$ imply that the Hamiltonian and momentum constraints, respectively reduce to
\begin{align}
\frac{3}{2}A_a^2 + \frac{2}{3}K^2 - 6H_0^2 &=0,\\
\partial_r A_a - \frac{2}{3}\partial_r K + 3\frac{A_a}{r} &=0.
\end{align}
Interestingly, upon setting $x=3A_a$, $y=2K$ these two equations can be rewritten as
\begin{align}
&x^2+y^2 = 36 \, H_0^2,\\
&\partial_rx-\partial_ry+3\frac{x}{r} =0,
\end{align}
the former being the implicit equation of a circle of radius $6H_0$. The general solution of these equations can be given in an implicit form in terms of a variable $\theta$ by defining $x=6H_0\cos\theta$, $y=6H_0\sin\theta$. One finds
\begin{equation}
-e^\theta\cos\theta = Cr^3,
\end{equation}
with $C$ an integration constant. The most trivial solution (and the only one in which the range of the coordinate radius $r$ is $[0,+\infty)$) involves setting $C=0$ and, therefore, $\cos\theta=0$. This corresponds to 
\begin{equation}
K = \pm 3H_0, \;\; A_a = 0.
\end{equation}
The minus sign is chosen in agreement with the cosmological expression for the background. 

The evolution of the gauge dynamics is performed in the harmonic gauge slicing in which the evolution equation for the lapse is
\begin{equation}
\partial_t\alpha = -\alpha^2K.
\label{eq:harmonic_slicing}
\end{equation}
This choice of gauge for the entire domain also fixes the gauge of the cosmological background dynamics. In addition to equations~(\ref{eq:Friedmann}) and~(\ref{eq:acc}), one thus also needs to solve 
\begin{equation}
\dot{\alpha}_{\tt bkg} = 3\alpha_{\tt bkg} \frac{\dot{a}}{a}.
\label{eq:lapsebkg}
\end{equation}
The only two independent variables for the background are $a$ and $\alpha_{\tt bkg}$. We choose to solve~(\ref{eq:acc}) and~(\ref{eq:lapsebkg}). Upon inserting~(\ref{eq:const_H0}), these equations become
\begin{align}
\dot{\alpha}_{\tt bkg} &= 3 \alpha_{\tt bkg}^2 H_0, \\
\frac{\ddot{a}}{a} &= 4\alpha_{\tt bkg}^2 H_0. 
\end{align}
We then use~(\ref{eq:const_H0}) to monitor the error on these in the manner of a constraint equation.

\subsection{Results}

We now come to discuss the stability of the scheme in the same terms as in Ref.~\cite{Montero2012}. This allows more straightforward comparisons. 

For values of the initial expansion factor of the order $H_0 \sim 10^{-3}$ or smaller, the exponential de Sitter expansion remains linear for time scales up to $t\sim 10$. In this case, the changes are small compared to the analysis carried out in Refs.~\cite{Montero2012,Alcubierre2011}. The code proceeds without difficulty yielding similar results for a value of the Courant-Friedrichs-Lewy (CFL) factor $\Delta t/\Delta r = 0.5$.

In what follows, we set $H_0 = 0.01$ and $\alpha_0=0.01$ that is, well within the exponential regime of the cosmological expansion.\footnote{In comparison, the value of $H_0$ for our Universe expressed in the units of this paper is of the order of $\sim10^{-23}$.} To proceed with such large values of the expansion, the CFL factor must be reduced. The results performed in this section have been obtained with $\Delta t/\Delta r = 0.25$.

\begin{figure}
  \begin{center}
    \includegraphics[width=0.5 \textwidth]{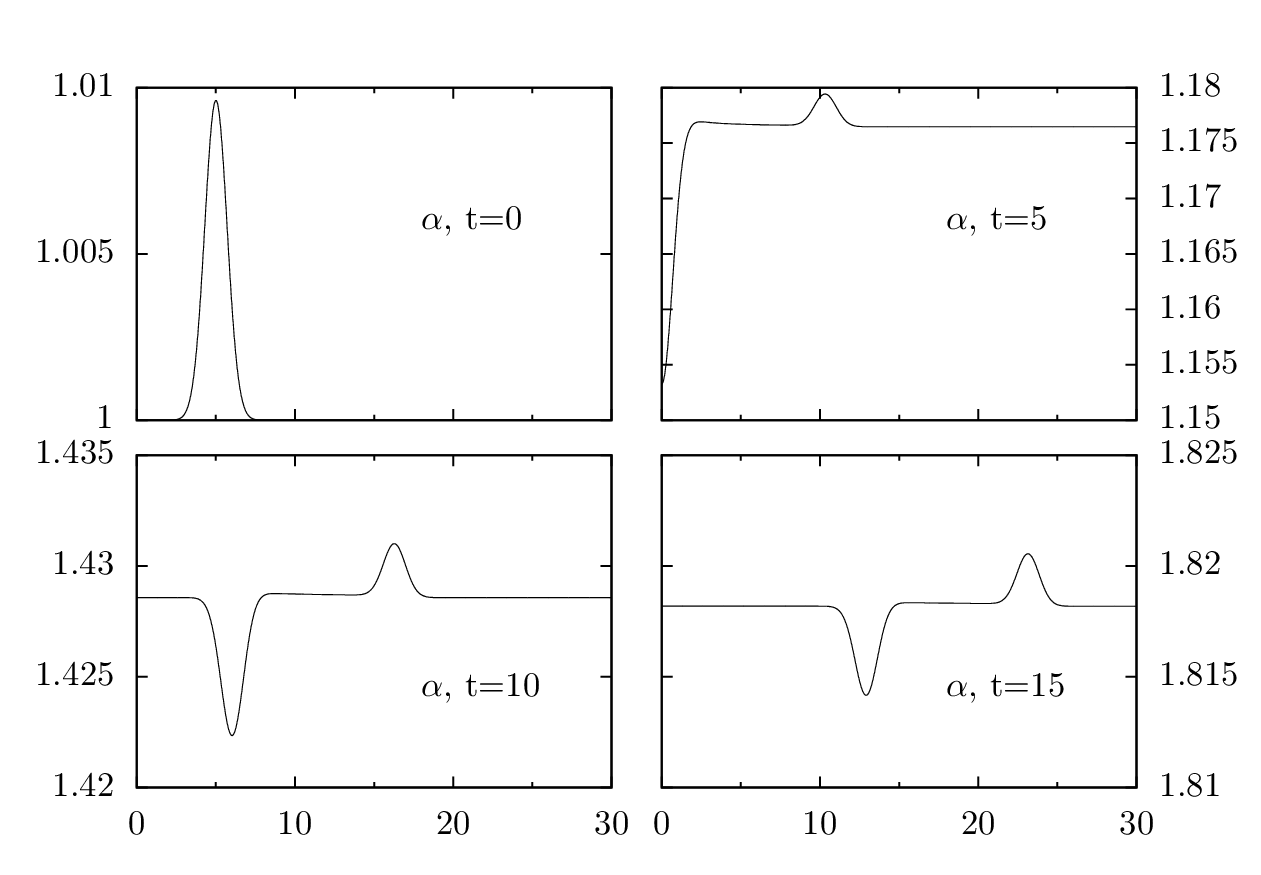}
  \end{center}
  \caption{Evolution of a pure gauge pulse in time on a de Sitter background with $H_0=0.01$. The asymptotic value of $\alpha$ gets rescaled during the evolution.}
  \label{fig:alpha_gd}
\end{figure}

\begin{figure}
  \begin{center}
    \includegraphics[width=0.5 \textwidth]{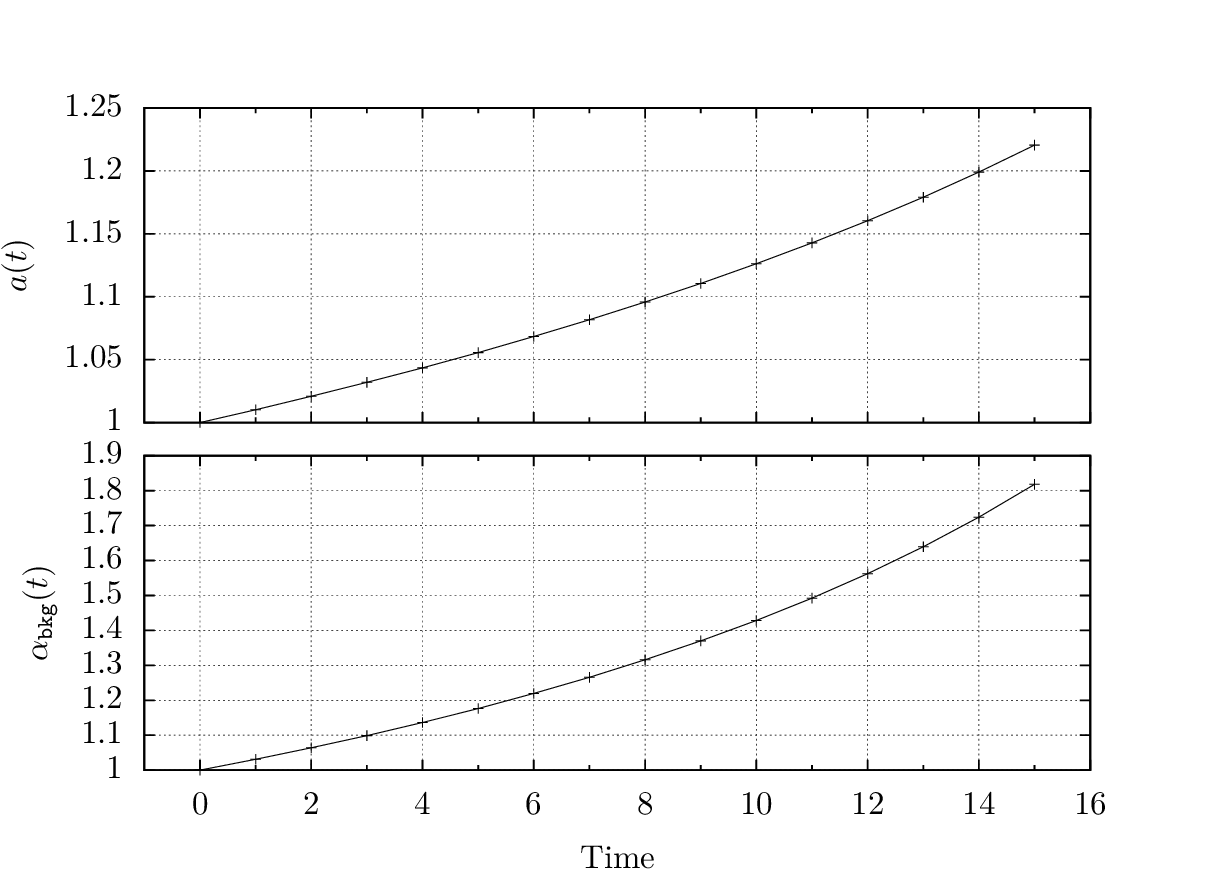}
  \end{center}
  \caption{Evolution of the scale factor $a(t)$ (upper panel) and background lapse function $\alpha_{\tt bkg}$ (lower panel).}
  \label{fig:cosmo_gd}
\end{figure}

The dynamics of the lapse in the harmonic slicing (\ref{eq:harmonic_slicing}) is that of a wave. As expected from the initial data, the initial gauge pulse splits in two parts travelling in opposite directions. One sees from Fig.~\ref{fig:alpha_gd}, which shows the radial profile of $\alpha$ for different values of the time, that the continuous background follows the evolution of $\alpha_{\tt bkg}$ imposed at the outer boundary condition and plotted in Fig.~\ref{fig:cosmo_gd} as a function of time.


\begin{figure}
  \begin{center}
    \includegraphics[width=0.5 \textwidth]{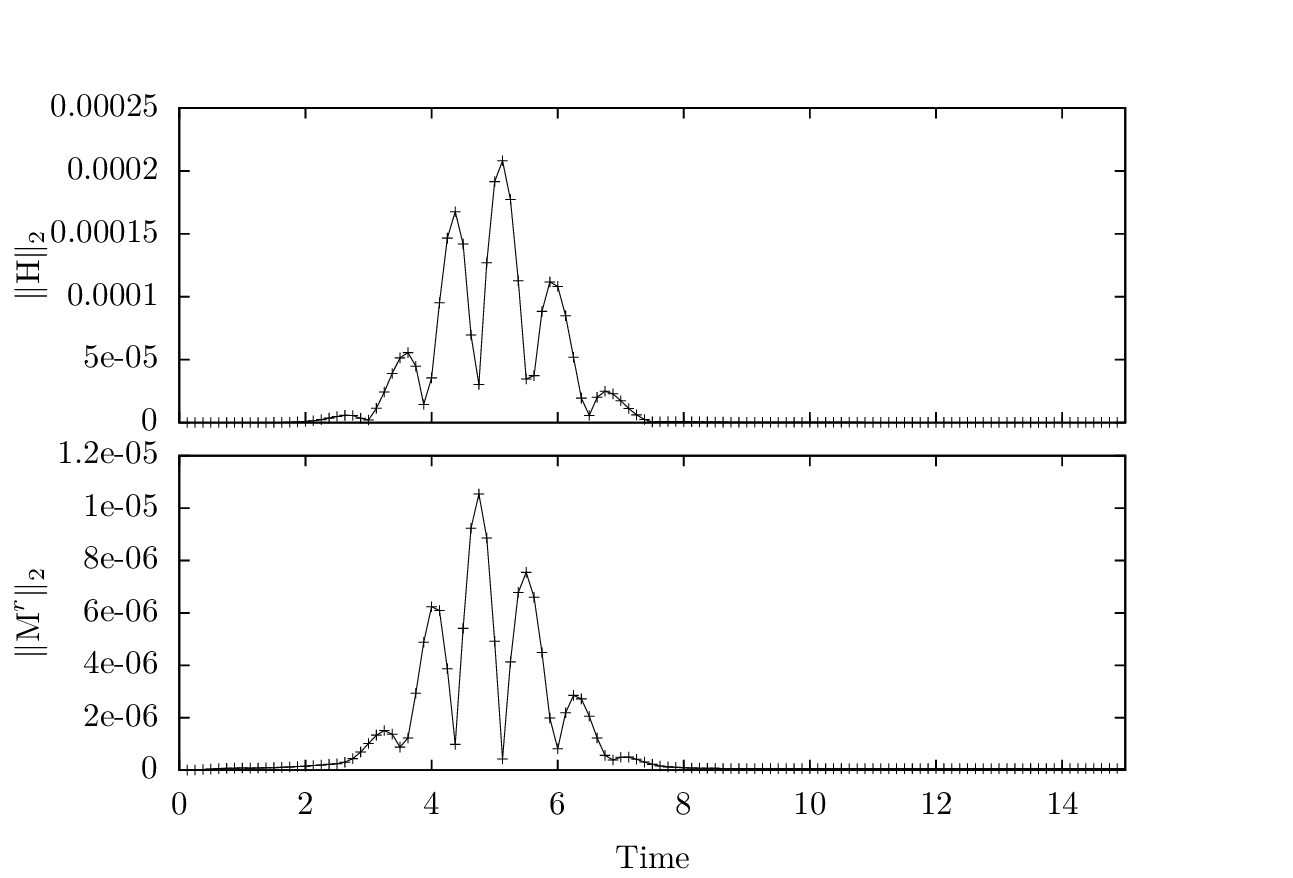}
  \end{center}
  \caption{$L^2$norm of the Hamiltonian (upper panel) and the momentum constraints (lower panel) in pure gauge dynamics ($\Delta r = 0.05$).}
  \label{fig:L2normHM}
\end{figure}

Figure.~\ref{fig:L2normHM} shows the $L^2$norm (root-mean-square) of the Hamiltonian and momentum constraints with a resolution of $\Delta r = 0.05$ as a function of time. The error hits a maximum when the left pulse hits the inner boundary. It goes down after it has bounced back and both pulses are travelling outward. 

\begin{figure}
  \begin{center}
    \includegraphics[width=0.5 \textwidth]{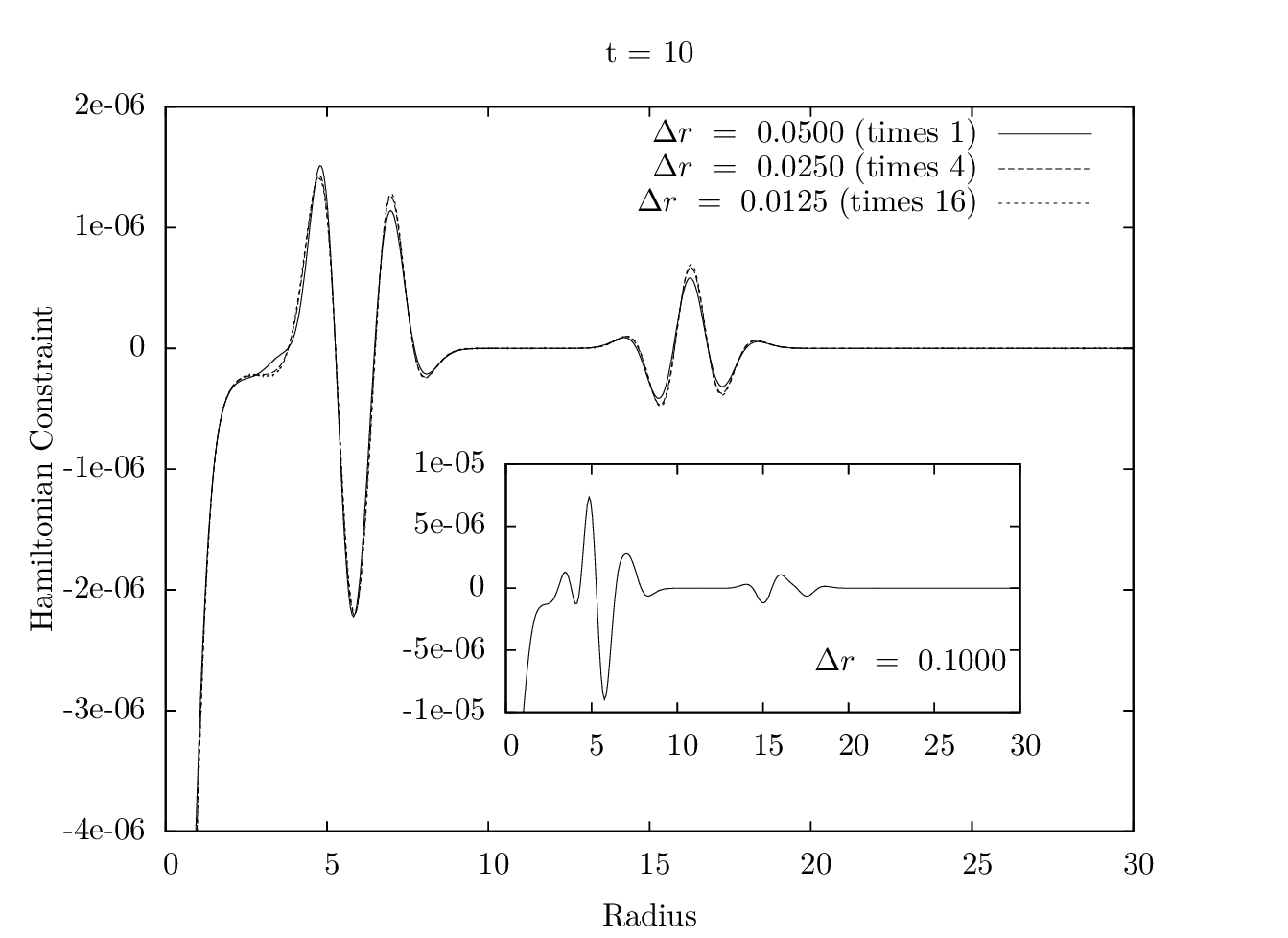}
  \end{center}
  \caption{Value of the Hamiltonian constraint in pure gauge dynamics for three different resolutions at $t=10$. The rescaling of the curves shows good agreement with the expected second-order convergence of the numerical method.}
  \label{fig:conv_gd}
\end{figure}

In Fig. ~\ref{fig:conv_gd}, we have plotted the Hamiltonian constraint for three values of the resolution. The rescaling of the error when resolution is doubled proves the good agreement with the expected second-order convergence of the numerical method. The curves shown here display a striking resemblance with the similar quantity obtained in Ref.~\cite{Montero2012} on a flat background. This implies that the expansion of the background leads only to small changes in the dynamics of the error. Aside for the fact that the CFL factor used here is smaller than $0.5$, another important difference with the simulations in~\cite{Montero2012} is that the convergence regime is attained for higher resolutions. This can be seen by looking at the Hamiltonian constraint profile shown in the inner plot of Fig.~\ref{fig:conv_gd}. The latter not being rescaled, its magnitude is of the correct order of magnitude but the profile is slightly different from those of the other curves.\footnote{Torres et al. \cite{Torres2014} have recently argued on how a higher value of the CFL factor can be used by substituting the harmonic slicing in favour of the Bona-Masso slicing with $f<1/3$. Such slicing is employed in the next section.}

\section{Application: Cosmological Spherical collapse}
\subsection{Lema\^itre-Tolman-Bondi solution}
We now apply the code to the study of the spherical collapse of pressureless matter (dust). This case is of practical interest in cosmology. It is usually studied using geodesic slicing gauge condition, $\alpha=1$. In this gauge, the most general solution to the Einstein's equations is the so-called Lema\^itre-Tolman-Bondi (LTB) solution. It can be summed up in the form of the metric line element
\begin{equation}
ds^2=-dt^2+\frac{a_\parallel^2(t,r)}{1+2 E_{\tt ltb}(r)}dr^2+a_\perp^2(t,r)r^2d\Omega^2,
\label{eq:LTBds2}
\end{equation}
with $a_\parallel=\partial_r (r \, a_\perp)$ and where $E_{\tt ltb}(r)$ is a free function.

The inhomogeneous counterparts of the Friedmann and acceleration equations are
\begin{align}
\frac{\dot{a}^2_\perp}{a^2_\perp} &= \frac{M(r)}{a^3_\perp} + \frac{2}{r^2}\frac{E_{\tt ltb}(r)}{a^2_\perp}, \label{eq:LTB1}\\
\frac{\ddot{a}_\perp}{a_\perp} &= -\frac{M(r)}{2a^3_\perp}\label{eq:LTB2},
\end{align}
in which $M$ is another free function related to the energy density through 
\begin{equation}
8\pi\rho = \frac{\partial_r(Mr^3)}{a_\parallel a^2_\perp r^2}.
\label{eq:rho_const_ltb}
\end{equation}
One of the main interests of using the geodesic slicing resides in the simplicity of the solution of the evolution equation for the dust density. Indeed, conservation of energy implies
\begin{equation}
\partial_t \rho+\frac{1}{2}(\gamma^{rr}\partial_t \gamma_{rr}+2\gamma^{\theta\theta}\partial_t \gamma_{\theta\theta})\rho = 0.
\label{eq:rhodot_dust}
\end{equation}
In terms of the LTB metric components, the solution of previous equation reads
\begin{equation}
\rho = \rho_0\frac{a_\parallel^0{a_\perp^0}^2}{a_\parallel a_\perp^2},
\end{equation}
where $\rho_0$ is the initial density profile. It can be shown that this is equivalent to Eq.~(\ref{eq:rho_const_ltb}).

\subsection{Initial data}
Building the initial data for the evolution of the LTB spacetime involves specifying an initial profile for three functions amongst $a_\perp$, $\dot{a}_\perp$, $E_{\tt ltb}$, $\rho$ and $M$. The remaining variables can then be inferred from Eqs.~(\ref{eq:LTB1}) and (\ref{eq:rho_const_ltb}). We wish to compare the evolution in the LTB and BSSN variables. We choose then to build the initial data from the constraints in the BSSN formulation and compute their equivalent in terms of the LTB variables.

To allow direct comparison, we choose as gauge variable $\alpha=1$. The initial values of the variables defined in Sec.~\ref{section:formalism} are
\begin{align}
\a(t=0) &= \b(t=0) = 1, \;\; K(t=0) = -3H_0,  \nonumber\\
A_a(t=0) &= A_b(t=0) = 0, \nonumber\\
E(r,t=0) &= [1+\delta_m(r)] \, \rho^0_{\tt bkg},
\label{eq:init_data_ltb}
\end{align}
where $\rho^0_{\tt bkg}=\rho_{\tt bkg}(t=0)$. In our work, we choose a density contrast profile in the form of a bump function,
\begin{equation}
\delta_m(r)=\delta_m^0\exp\left(-\frac{r^2}{r_0^2-r^2}\right),
\label{eq:bump}
\end{equation}
where $\delta_m^0$ and $r_0>0$ are constants. This profile has the property of being smooth and has a compact support spanning the region $[0,r_0]$. Other profiles have been used with similar success though not documented here.

In particular, the choice of $K$ and $A_a$ imposes that
\begin{equation}
H_0 = \frac{\dot{a}_0}{a_0} = \gamma^{rr} \partial_t \gamma_{rr}|_{t=0} = \gamma^{\theta\theta} \partial_t \gamma_{\theta\theta}|_{t=0}.
\label{eq:initH}
\end{equation}
The equation for the initial value of the conformal factor is found by plugging initial conditions from Eq.~(\ref{eq:init_data_ltb}) into the Hamiltonian constraint, which reduces to
\begin{equation}
a^{-2}\psi^{-5}\left(\partial_r^2\psi+\frac{2}{r}\partial_r\psi\right)+6H_0^2=16\pi\rho_{\tt bkg}^0(1+\delta_m^0(r)).
\end{equation}
Using the Friedmann equation, the previous expression becomes
\begin{equation}
\partial_r^2\psi+\frac{2}{r}\partial_r\psi = 16\pi \, \rho_{\tt bkg}^0 \, \delta_m^0(r) \, a_0^{2} \, \psi^{5}.
\end{equation}
Following Ref.~\cite{Shibata1999}, this equation is solved numerically as a boundary value problem with conditions 
\begin{align}
\partial_r\psi\rightarrow &~0, & \text{for} ~r&\rightarrow 0;\\
\psi\rightarrow &~1+\frac{C_\psi}{2r}, & \text{for} ~r&\rightarrow\infty.
\end{align}
The parameter $C_\psi$ is adjusted by specifying an additional outer boundary condition: 
\begin{equation}
\partial_r\psi \rightarrow \frac{-C_\psi}{2r^2}. 
\end{equation}

\begin{figure}
  \begin{center}
    \includegraphics[width=0.5 \textwidth]{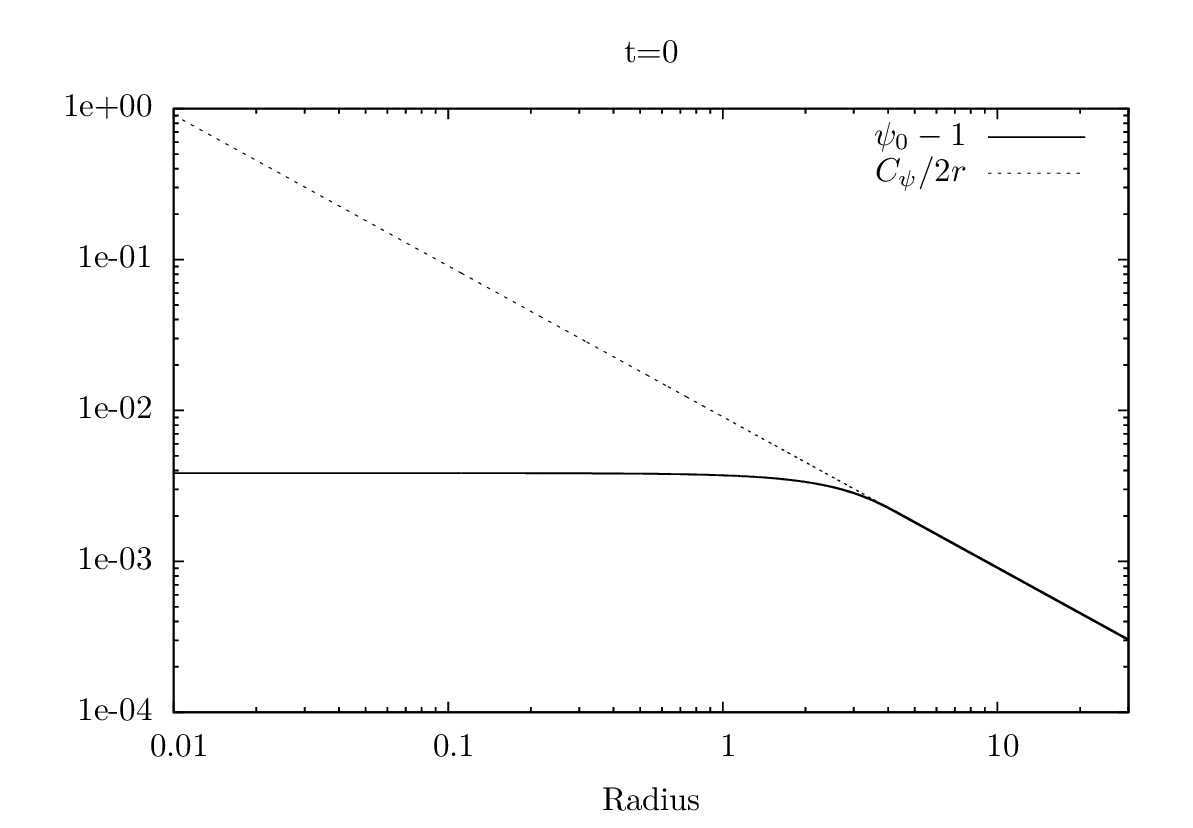}
  \end{center}
  \caption{Initial conformal factor in the case of a dust matter overdensity of central value $\delta_m^0=0.1$ (plain line). The solution agrees well with the asymptotic value imposed as boundary condition (dotted line).}
  \label{fig:psi0}
\end{figure}

The solution to the initial boundary value problem is shown in Fig.~\ref{fig:psi0} for $\delta_m^0 = 0.1$ and $r_0=5$ (plain line). Its behaviour agrees well with the imposed asymptotic solution (dashed line).

We now come to set the initial data in terms of the LTB variables. Since the analysis is performed in the zero-shift gauge, it can be assumed that the radius coordinates of both ansatz of the metric should only differ up to a constant factor throughout the integration. Setting this factor to 1 in the initial data allows to compare the metric components themselves between both methods. 

From the decomposition introduced in Eq.~(\ref{eq:ansatz}) and using the fact that $a_\parallel=\partial_r (r \, a_\perp)$, one obtains the initial values of $a_\perp$ and $a_\parallel$ by differentiation:
\begin{align}
a^0_\perp&=\psi_0^2 \, a_0, \\
a^0_\parallel&=\psi_0^2 \, a_0 + 2\psi_0 \, \frac{d\psi_0}{dr} \, a_0 \, r.
\end{align}
By comparison of the radial part of the spatial metric in both gauge, one then finds the form of the energy function $E_{\tt ltb}(r)$ of (\ref{eq:LTBds2}). In agreement with~(\ref{eq:initH}), the initial time derivatives are
\begin{equation}
\dot{a}^0_\perp = a^0_\perp H_0, \;\; \dot{a}^0_\parallel = a^0_\parallel H_0.
\end{equation}
Using Eq.~(\ref{eq:LTB1}), $M(r)$ is deduced and can then be used for the evolution of $\dot{a}$ using Eq.~(\ref{eq:LTB2}). 

\subsection{Evolution and results in geodesic slicing ($\alpha=1$)}

In terms of the BSSN variables, taking into account that $\a\,\b^2=1$ always holds, the solution of Eq.~(\ref{eq:rhodot_dust}) is
\begin{equation}
\rho=\rho_0\left(\frac{a_0^3 \, \psi_0^6}{a^3 \, \psi^6} \right),
\label{eq:rescale_BSSN}
\end{equation}
where $\psi_0 = \psi (t=0)$. This expression generalizes the rescaling equation of dust in cosmology to the case of a nonhomogeneous spacetime.

The background evolution proceeds in the same way as for the case of gauge dynamics. We solve the acceleration equation, which in the geodesic slicing and in presence of dust only reduces to 
\begin{equation}
\frac{\addot}{a} = -\frac{8\pi}{6}\rho_{\tt bkg}.
\label{eq:acc_dust}
\end{equation}
The homogeneous part of the dust energy density is evolved simply as $\rho_{\tt bkg} = \rho_{\tt bkg}^0a_0^3/a^3$.

\begin{figure}
  \begin{center}
    \includegraphics[width=0.5 \textwidth]{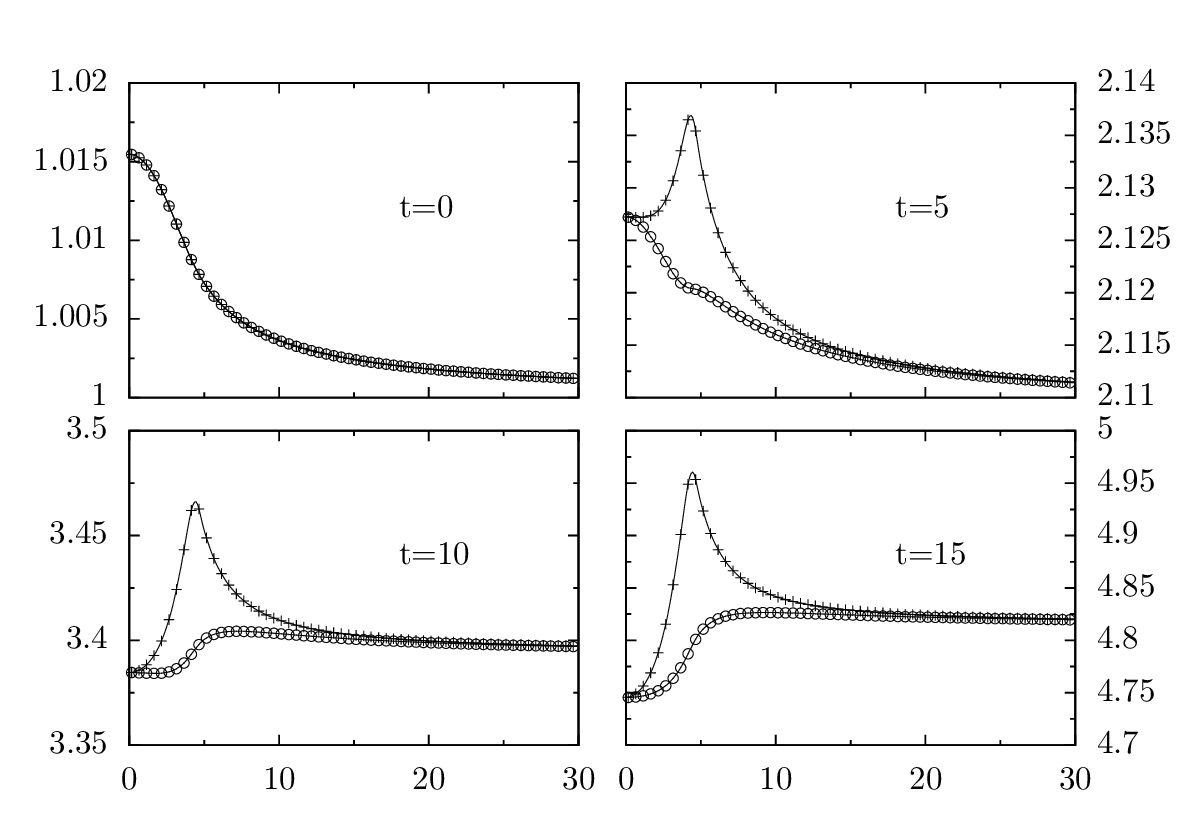}
  \end{center}
  \caption{Metric components $\gamma_{rr}$ (top curves) and $\gamma_{\theta\theta}/r^2$ (bottom curves). The plain lines show the result of the evolution of the LTB variables, while crosses and circles are the evolution of the BSSN equations. The curves coincide at initial time (upper-left panel). The maximum of the relative difference between the curves is of the order $\sim 10^{-5}$.}
  \label{fig:compLTB}
\end{figure}

Figure.~\ref{fig:compLTB} shows the result of the evolution of the $\gamma_{rr}$ and $\gamma_{\theta\theta}$ 3-metric components using the LTB variables (lines) and the BSSN equations (crosses and circles) for different values of the coordinate time $t$, up to $t=15$. The shape of the curves remains basically unchanged for subsequent values of $t$. The simulation has been performed using $H_0=0.1$ and $\Delta r=0.1$. The maximum of the relative difference between the analytical and numerical values for both metric components shown in Fig.~\ref{fig:compLTB} is of the order $\sim10^{-5}$ and is lower with higher resolution. These simulations were performed using a CFL factor equal to $0.5$.

\begin{figure}
  \begin{center}
    \includegraphics[width=0.5 \textwidth]{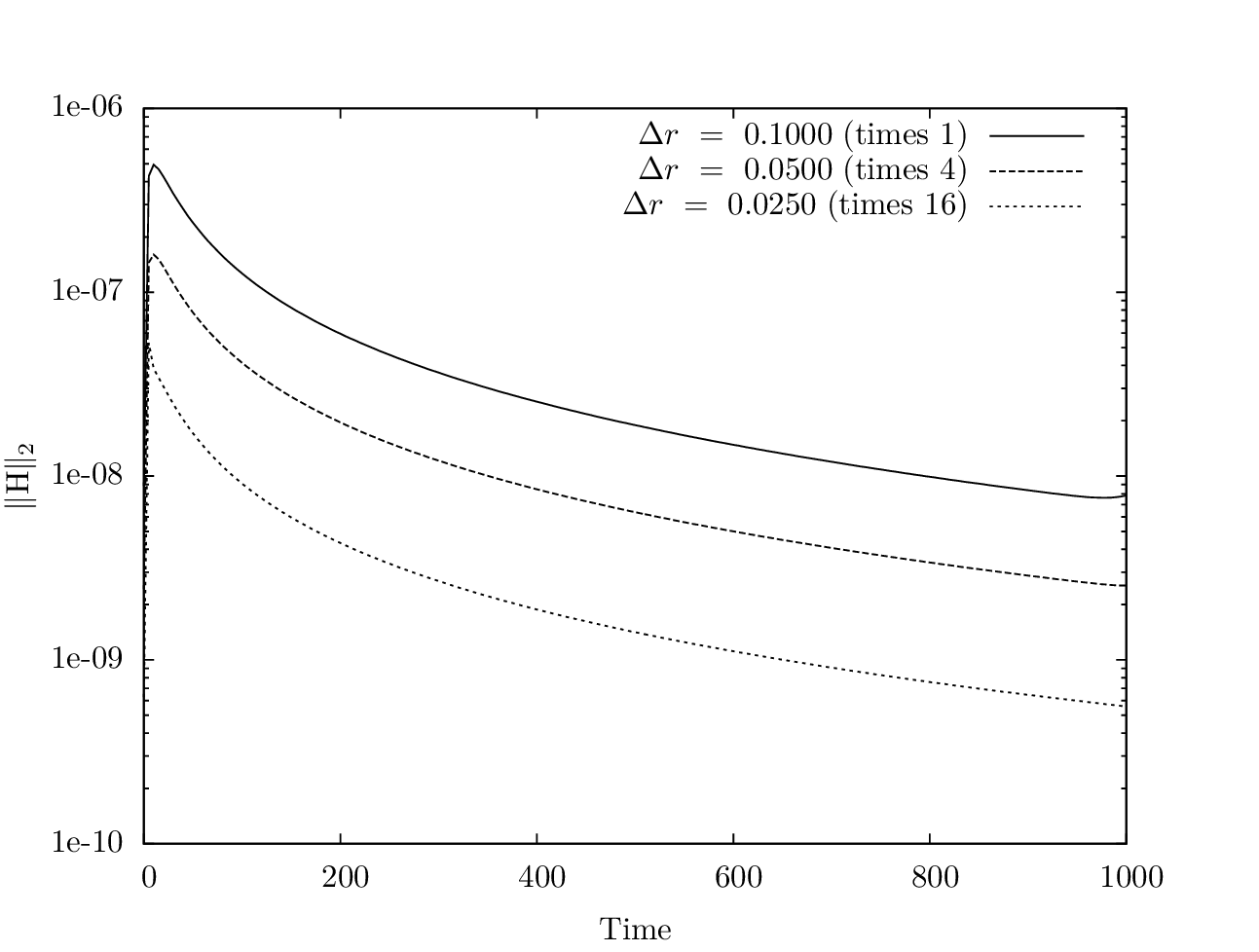}
  \end{center}
  \caption{$L^2$norm of the Hamiltonian constraint for long-time evolution of the collapse of dust for different resolutions.}
  \label{fig:L2_dust}
\end{figure}

The long-term stability analysis of the code is better analyzed by looking at the evolution of the $L^2$-norm of the Hamiltonian constraint, displayed in Fig.~\ref{fig:L2_dust} for different resolutions. We obtain similar shapes in all the curves. The difference in magnitude despite the rescaling indicates an order of convergence above second order. We accept this to be a result of the fact that the evolution of matter is simple enough to make the dominant error come from the finite difference scheme used to compute spatial derivatives (fourth order) rather than the time evolution integration. 

\begin{figure}
  \begin{center}
    \includegraphics[width=0.5 \textwidth]{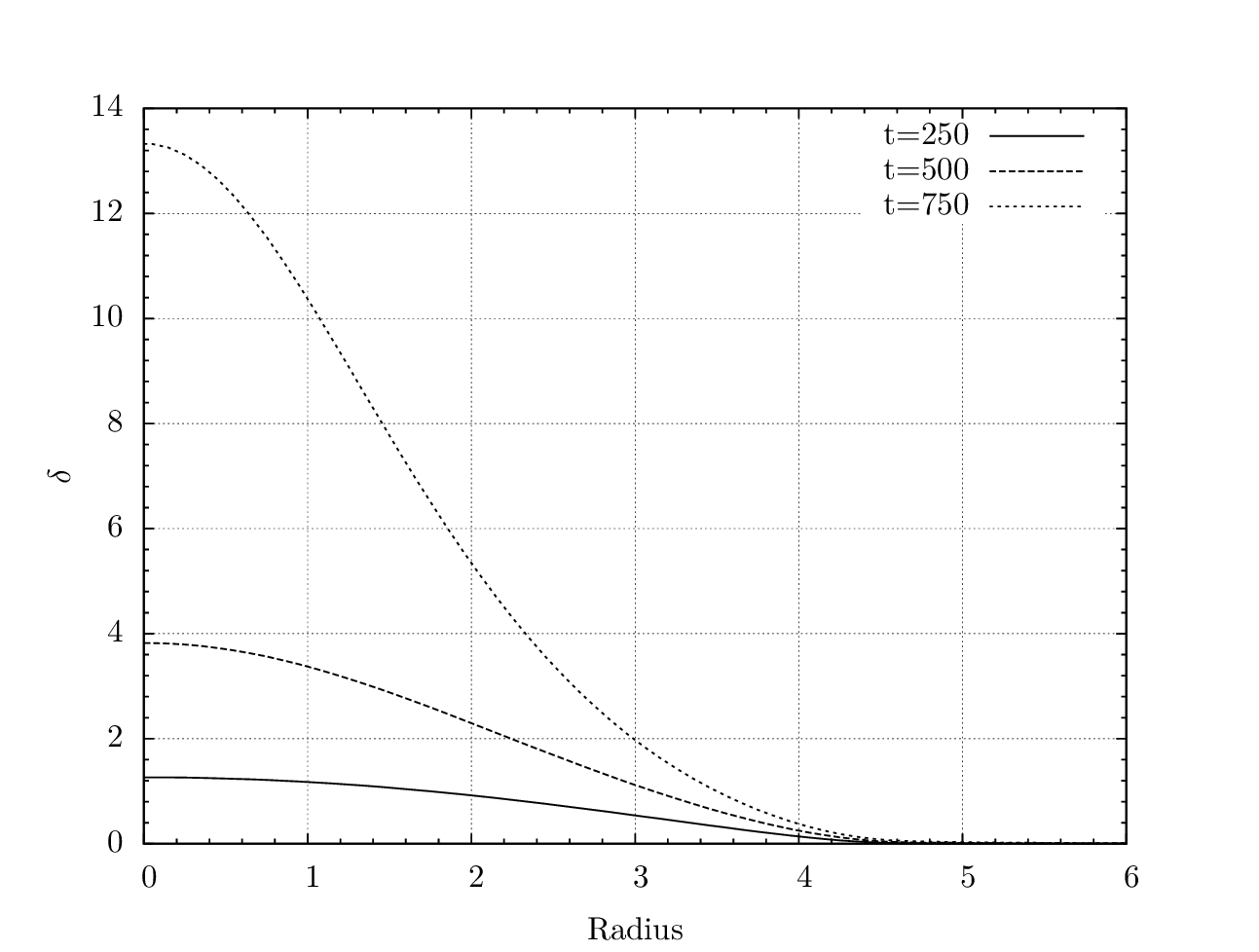}
  \end{center}
  \caption{Evolution of the dust density contrast profile.}
  \label{fig:delta(t,r)}
\end{figure}

\begin{figure}
  \begin{center}
    \includegraphics[width=0.5 \textwidth]{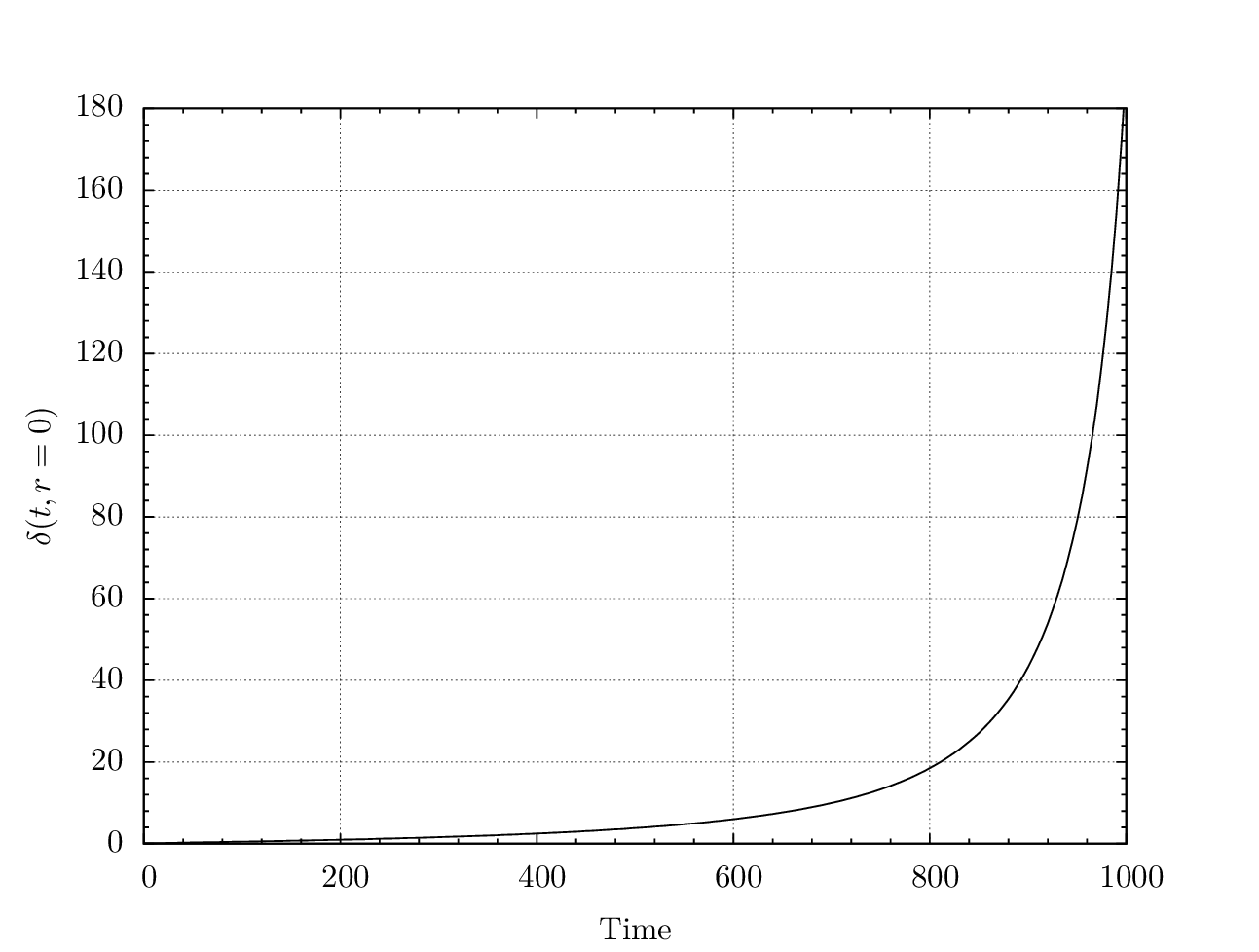}
  \end{center}
  \caption{Long-term evolution of the central overdensity of dust matter $\delta_c:=\delta(t,r=0)$.}
  \label{fig:delta_cosmo}
\end{figure}

\begin{figure}
  \begin{center}
    \includegraphics[width=0.5 \textwidth]{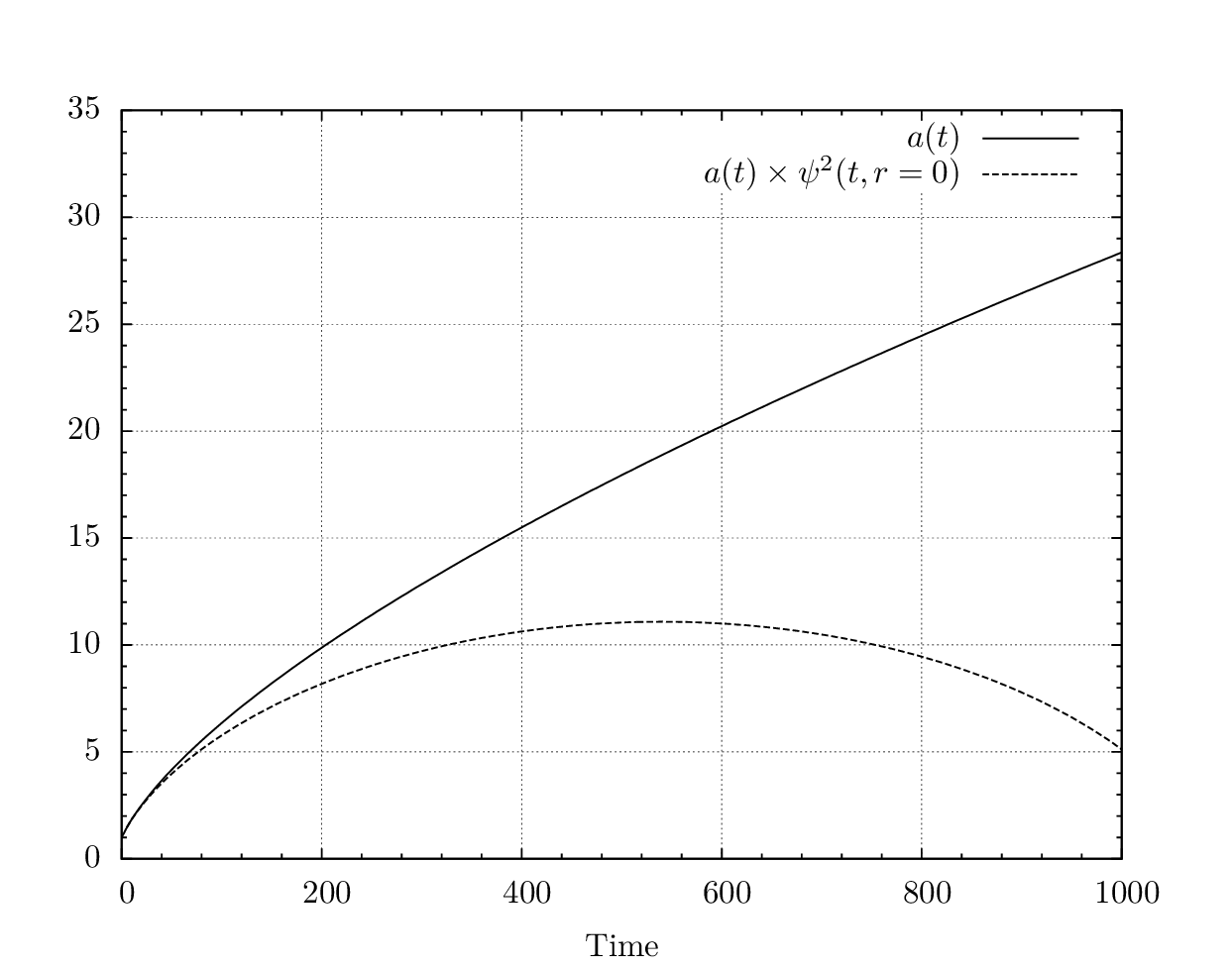}
  \end{center}
  \caption{Evolution of the background scale factor in time (plain line) compared with the central conformal expansion factor (dotted line) defined as $a(t) \, \psi^2(t,r=0)$.}
  \label{fig:cosmo_dust}
\end{figure}

The dust density contrast profile, $\delta(t,r)$, is defined through the expression $\rho(t,r)=\rho_{\tt bkg}(t)(1+\delta(t,r))$. It is plotted for three different values of $t$ in Fig.~\ref{fig:delta(t,r)}. The profile grows exponentially and its shape changes in time departing from the initial bump profile given in Eq.~(\ref{eq:bump}). We see no effect of the central coordinate singularity on the profile. This shows the good reliability of the PIRK algorithm. One useful tool in cosmology is the value of the central density contrast as a function of time, plotted in Fig.~\ref{fig:delta_cosmo}. The numerical simulation can be used to investigate the nonlinear regime of growth of dust matter density. The background scale factor is shown in Fig.~\ref{fig:cosmo_dust}, along with the local conformal scale factor defined as the product $a^2(t) \, \psi(t,r=0)$. In ordinary studies, it is assumed that virialisation should occur when the local scale factor decreases to half its maximum value~\cite{Padmanabhan1993}.  


\subsection{Solution with a dynamical lapse ($\alpha\neq1$)}
\label{sect:dust_alpha}

The study performed in the previous section can be modified to accommodate the case where the lapse is dynamical. Equation~(\ref{eq:rescale_BSSN}) is only valid in geodesic slicing. 

The hydrodynamic equation for the evolution of matter, which can be derived from the local conservation of baryon number and energy-momentum, can be written as a first-order hyperbolic system of conserved variables, known as the Valencia formulation~\cite{Ibanez1999}. This formulation ensures very good stability of the matter evolution. The corresponding expression in spherical symmetry is  
\begin{equation}
\label{cons.eq.}
	\partial_{t}{\bf{U}}+\partial_{r}{\bf{F}}^{r}={\bf{S}},
\end{equation}
where ${\bf{U}}=\sqrt{\gamma}(D,S_{r},\tau)$ is the vector of conserved variables, being 
\begin{align}
	D & = \rho W,\\
	S_r & = \rho h W^2v_r,\\
	\tau & = \rho h W^2 - p - D.
\end{align}
$W$ is the Lorentz factor $W:=(1-v_rv^r)^{-\frac{1}{2}}$, $v_r$ the speed of the fluid relative to the Eulerian observer and $h$ is the enthalpy of the fluid.
The components of ${\bf{F}}^r$ and ${\bf{S}}$ are, respectively, the fluxes and source functions. Their explicit expressions can be found in Appendix~\ref{sect:Hydro}. 

\begin{figure}
  \begin{center}
    \includegraphics[width=0.5 \textwidth]{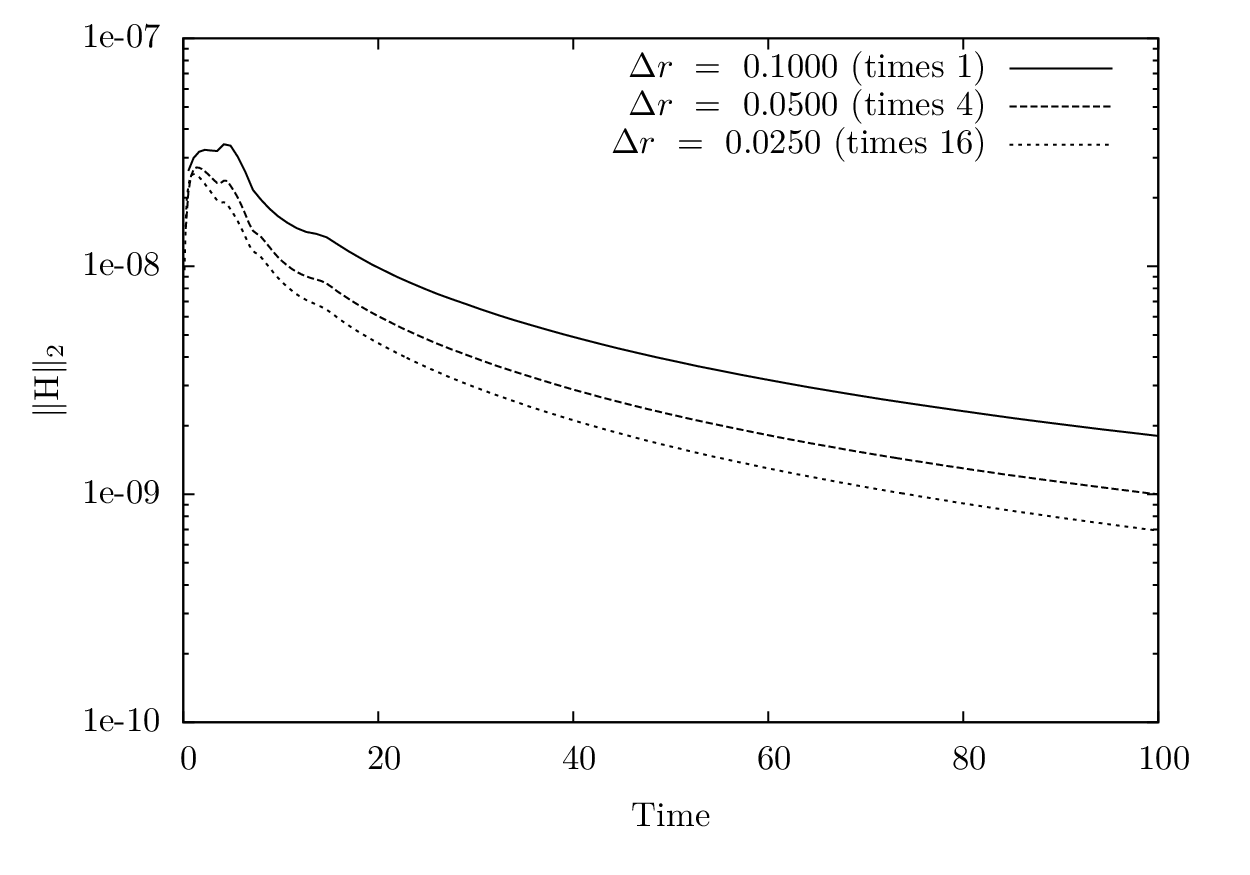}
  \end{center}
  \caption{$L^2$norm of the Hamiltonian constraint as a function of the cosmic time for the evolution of dust using the Bona-Masso slicing with $f=0.333$ for different resolutions.}
  \label{fig:L2_H_alpha}
\end{figure}

\begin{figure}
  \begin{center}
    \includegraphics[width=0.5 \textwidth]{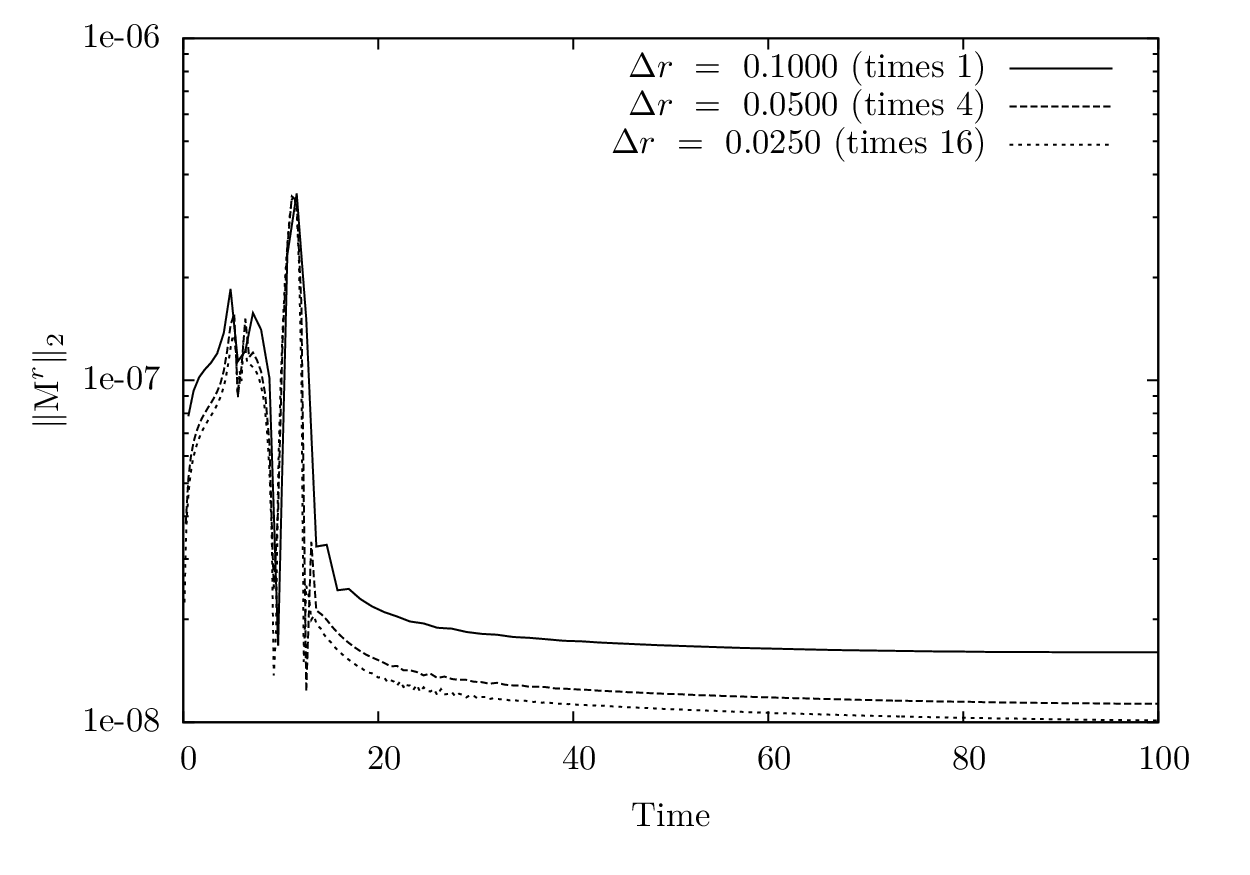}
  \end{center}
  \caption{$L^2$norm of the momentum constraint as a function of the cosmic time for the evolution of dust using the Bona-Masso slicing with $f=0.333$ for different resolutions.}
  \label{fig:L2_M_alpha}
\end{figure}

Following Ref.~\cite{Torres2014} we evolve the lapse using $\partial_t\alpha=-\alpha^2fK$, with $f=0.333$. The cosmic time $t_\text{cos}$ is related to the computational time $t$ through $dt_\text{cos}=\alpha_{\tt bkg}dt$. As the lapse grows monotonically in our case, so does the cosmic time interval. The Bona-Masso slicing appears as a poor choice for long evolution in the case where the expansion factor is so big in magnitude. It is, however, very interesting in order to prove the stability of the method. We use a HLLE solver and MC slope limiter to go from conserved to primitive variables. Figure.~\ref{fig:L2_H_alpha} and Fig.~\ref{fig:L2_M_alpha} show the evolution of the Hamiltonian and momentum constraints for different values of the resolution as a function of the cosmic time. We have obtained the second order of convergence at the start of the simulation and even higher orders at later times. We have used $\Delta t/\Delta r = 0.5$. The maximum of the error in the momentum constraint appears in the range corresponding to a change in the shape of the metric components much in the same way as what is shown in Fig.~\ref{fig:compLTB} in the case of geodesic slicing.


\section{Conclusion and Perspectives}
\label{section:conclusion}
This work is a first step towards the study of simulations of nonlinear structure formations in the presence of exotic varieties of matter or the further study of the formation of PBH already engaged in Ref.~\cite{Shibata1999}. Further directions include the application study of the cosmological spherical collapse including new scalar degrees of freedom (quintessence). This would provide an invaluable tool to discriminate between various dark energy candidates. Such study is currently investigated by the authors. The PIRK methods are a very good tool, compared to explicit RK methods, to undertake such study in the BSSN formalism as these are especially suited to the study of wavelike equations in spherical coordinates. The approach presented here can also be straightforwardly adapted to the case of a fluid with pressure. Another interesting application to derive from the full relativistic computation of the metric variables is to analyse the geodesics around the structure as it collapses. The method presented here can be used for that purpose if teamed with a geodesic dedicated code such as the \texttt{GYOTO} code presented in \cite{Vincent2011}.

We have presented a full relativistic numerical method suited for cosmological studies of problems with spherical symmetry. The stability of the algorithm at the center of coordinates is ensured by the use of the PIRK methods. From this point of view, the present paper generalizes the results obtained in Ref.~\cite{Montero2012}, in which the same methods were applied to asymptotically flat spacetimes. We have given a generalization of the treatment of a radiative boundary condition to the case of a dynamical background and provided proofs of the stability and convergence of the code by solving for the dynamics of a pure gauge pulse on an expanding de Sitter background. One of the key steps in the process of building a numerical scheme on a flat background involves testing it on the most basic spherically symmetric vacuum solution, namely the Schwarzschild black hole. We have generalized this study by applying our code to study the numerical spherical collapse of dust which is adequately described by the LTB solution. We have shown how our code reproduces the same solution in presence of identical initial data and by comparing the metric components and we have demonstrated its stability in the case where the lapse function is dynamical.

\begin{acknowledgments}
The authors warmly thank Dr.~P.~Montero for useful discussions and his practical expertise regarding the numerical methods used in this work. J.~R. is supported by a FRS-FNRS (Belgian Fund for Scientific Research) Research Fellowship. I.~C.-C. and \& A.~F. are partially supported by the ARC convention No. 11/15-040. This work was also supported by SN2NS Project No. ANR-10-BLAN-0502, the Fonds de la Recherche Scientifique - FNRS under Grant No. 4.4501.05, ERC Starting Grant No. CAMAP-259276, and the Grants No. AYA2010-21097-C03-01 and AYA2013-40979-P of the Spanish MICINN. Computations were performed at the ``plate-forme technologique en calcul intensif'' (PTCI) of the University of Namur, Belgium, with the financial support of the FRS- FNRS (conventions No. 2.4617.07. and No. 2.5020.11). 
\end{acknowledgments}

\appendix
\section{Algebraic expressions of used quantities}
\label{sect:algexpr}
The following expressions are extracted from Ref.~\cite{Alcubierre2011}. They are repeated here for the convenience of the reader with slight modifications regarding the change of notation, mainly the introduction of the scale factor in the 3-metric~:

\begin{align}
R^r_r &= -\frac{1}{a\a\psi}\left[\frac{\partial_r^2\a}{2\a}-\a\partial_r\hat{\Delta}^r-\frac{3}{4}\left(\frac{\partial_r\a}{\a}\right)^2+\frac{1}{2}\left(\frac{\partial_r\b}{\b}\right)^2\right.\nonumber\\
	 &\left.-\frac{1}{2}\hat{\Delta}^r\partial_r\a+\frac{\partial_r\a}{r\b}+\frac{2}{r^2}\left(1-\frac{\a}{\b}\right)\left(1+\frac{r\partial_r\b}{\b}\right)\right.\nonumber\\
	 &\left.+4\frac{\partial^2_r\psi}{\psi}-4\left(\frac{\partial_r\psi}{\psi}\right)^2-2\left(\frac{\partial_r\psi}{\psi}\right)\left(\frac{\partial_r\a}{\a}-\frac{\partial_r\b}{\b}-\frac{2}{r}\right)\right].
\end{align}
\begin{align}
R &= -\frac{1}{a\a\psi}\left[\frac{\partial_r^2\a}{2\a}+\frac{\partial_r^2\b}{\b}-\a\partial_r\hat{\Delta}^r-\left(\frac{\partial_r\a}{\a}\right)^2\right.\nonumber\\
	 &\left.+\frac{1}{2}\left(\frac{\partial_r\b}{\b}\right)^2+\frac{2\partial_r\b}{r\b}\left(3-\frac{\a}{\b}\right)+\frac{4}{r^2}\left(1-\frac{\a}{\b}\right)\right.\nonumber\\
	 &\left.+8\frac{\partial^2_r\psi}{\psi}-8\left(\frac{\partial_r\psi}{\psi}\right)\left(\frac{\partial_r\a}{2\a}-\frac{\partial_r\b}{\b}-\frac{2}{r}\right)\right].
\end{align}
\begin{align}
\nabla^2\alpha &= \frac{1}{a\a\psi}\left[\partial_r^2\alpha-\partial_r\alpha\left(\frac{\partial_r\a}{2\a}-\frac{\partial_r\b}{\b}-2\frac{\partial_r\psi}{\psi}-\frac{2}{r}\right)\right].\\
\nabla^r\nabla_r\alpha &= \frac{1}{a\a\psi}\left[\partial_r^2\alpha-\partial_r\alpha\left(\frac{\partial_r\a}{2\a}+2\frac{\partial_r\psi}{\psi}\right)\right].
\end{align}
\vspace{0cm}

\section{Definition of the hydrodynamical variables}
\label{sect:Hydro}
The following formulae are the expressions for the fluxes and source terms used in Sec.~\ref{sect:dust_alpha}. These are extracted from Ref.~\cite{Montero2012} and given for arbitrary lapse and shift~:

\begin{align}
{\bf{F}}^{r} & =\sqrt{-g}\left[D(v^{r}
  -\beta^{r}/\alpha), \right. \nonumber \\
&  \left. S_{r}(v^{r}-\beta^{r}/\alpha)+p, \right. \nonumber \\
& \left. \tau  (v^{r}-\beta^{r}/\alpha)+p v^r\right],
\end{align}

\begin{align}
	{\bf{S}} &= \sqrt{-g}\left[0, 
T^{00}\left(\frac{1}{2}(\beta^{r})^{2}\partial_{r}\gamma_{rr}
- \alpha\partial_{r}\alpha \right) \right. \nonumber\\
 & \left. + T^{0r}\beta^r\partial_{r}\gamma_{rr}+T^{0}_{r}\partial_{r}\beta^{r}
  +\frac{1}{2}T^{rr}\partial_{r}\gamma_{rr}, \right. \nonumber\\ 
 & (T^{00}\beta^{r}+T^{0r})(\beta^{r}K_{rr} - \partial_{r}\alpha)+T^{rr}K_{rr}
  \bigg].
\end{align}

\bibliographystyle{apsrev}
\bibliography{bibfile}

\end{document}